\documentclass{article}
\usepackage{spconf,amsmath,graphicx}
\usepackage{cite}
\usepackage{amssymb}
\usepackage{mathrsfs}
\usepackage{amsmath}
\usepackage{bm}
\usepackage{url}
\usepackage{booktabs}
\usepackage{multirow}
\usepackage{enumitem}
\usepackage{caption}

\title{
6D\lowercase{o}F SELD: Sound Event Localization and Detection Using\\ Microphones and Motion Tracking Sensors on self-motioning human
}

\name{
Masahiro Yasuda,
Shoichiro Saito,
Akira Nakayama,
Noboru Harada
}
\vspace{-10pt}
\address{NTT Corporation, Japan
\vspace{-10pt}
}

\begin{document}
\ninept
\maketitle

\begin{abstract}
We aim to perform sound event localization and detection (SELD) using wearable equipment for a moving human, such as a pedestrian. 
Conventional SELD tasks have dealt only with microphone arrays located in static positions. 
However, self-motion with three rotational and three translational degrees of freedom (6DoF) shall be considered for wearable microphone arrays. 
A system trained only with a dataset using microphone arrays in a fixed position would be unable to adapt to the fast relative motion of sound events associated with self-motion, resulting in the degradation of SELD performance. 
To address this, we designed \textit{6DoF SELD Dataset}\footnote{\scriptsize{The dataset is available in \url{https://github.com/nttrd-mdlab/6dof-seld}\quad (DOI: 10.5281/zenodo.10473531)}} for wearable systems, the first SELD dataset considering the self-motion of microphones. 
Furthermore, we proposed a multi-modal SELD system that jointly utilizes audio and motion tracking sensor signals.
These sensor signals are expected to help the system find useful acoustic cues for SELD on the basis of the current self-motion state. 
Experimental results on our dataset show that the proposed method effectively improves SELD performance with a mechanism to extract acoustic features conditioned by sensor signals.

\end{abstract}

\begin{keywords}
sound event localization and detection, motion tracker, six degrees of freedom, microphone array, dataset
\end{keywords}
\vspace{-4pt}
\section{Introduction}
\label{sec:intro}
\vspace{-4pt}

Sound event localization and detection (SELD) is a combined task of sound event detection, which estimates the class of event and its onset/offset time, and sound source localization~\cite{SELD, SELD2023}. 
In this study, we newly defined and addressed 6DoF SELD, a SELD using microphone arrays worn by a self-moving human in six degrees of freedom (6DoF).
Here, 6DoF is the sum of three rotational and three translational degrees of freedom, corresponding to behaviors such as walking, looking around, and bending over.
The output of 6DoF SELD is similar to that of SELD, but the direction of arrival (DOA) of the sound source is estimated in relative coordinates for the head's orientation. For example, when the sound source is fixed, the estimated DOA moves in the opposite direction of the head motion.
One promising application is pedestrian safety assistance through notification of approaching vehicles and humans. Another application is in immersive communication, where the status of the surrounding environment is shared remotely~\cite{immersive1, immersive2}. 
Moreover, SELD on moving vehicles, such as autonomous cars~\cite{SmartCar1,SmartCar2} and surveillance drones~\cite{drone}, can also be considered an application of 6DoF SELD. 
As a practical constraint in these applications, the system shall be developed under the causal constraint of being able to operate online~\cite{OnlineSELD}.

\begin{figure}[h!]
\centering
\includegraphics[width=\linewidth]{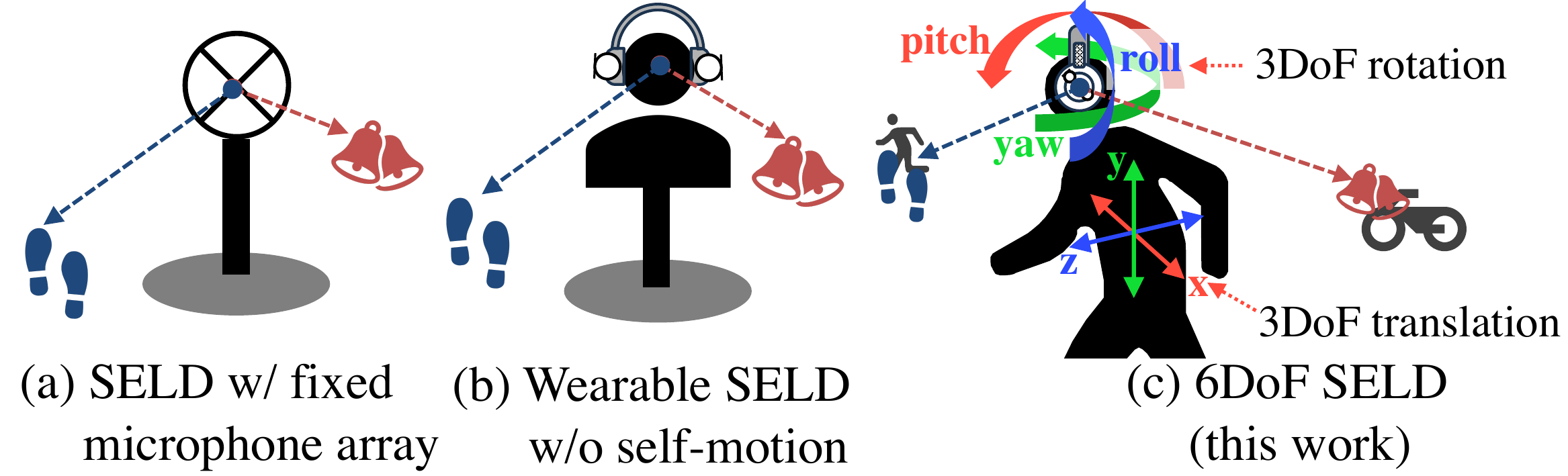}
  \vspace{-15pt}
  \caption{Conventional and our problem settings of SELD}
  \label{fig:overview}
  \vspace{-15pt}
\end{figure}

Conventional SELD systems mainly use first-order ambisonics (FOA) signals as input and the deep neural networks (DNN) as a regression or classification function for SELD, as shown in Fig.~\ref{fig:overview} (a), (b)~\cite{dataset2019,dataset2020,dataset2021,dcase2023baseline,dcase2023sota, WearableSELD}.
In particular, Fig.~\ref{fig:overview} (a) shows the most investigated SELD systems that use FOA signals. Such a system would be useful in applications where a device placed in a room, such as a smart speaker, analyzes indoor events. Although the SELD using wearable microphone arrays has rarely been addressed, our previous work~\cite{WearableSELD} addressed wearable SELD by using a microphone array attached to a head and torso simulator (HATS).

\begin{figure*}[t]
\centering
\includegraphics[width=0.96\linewidth]{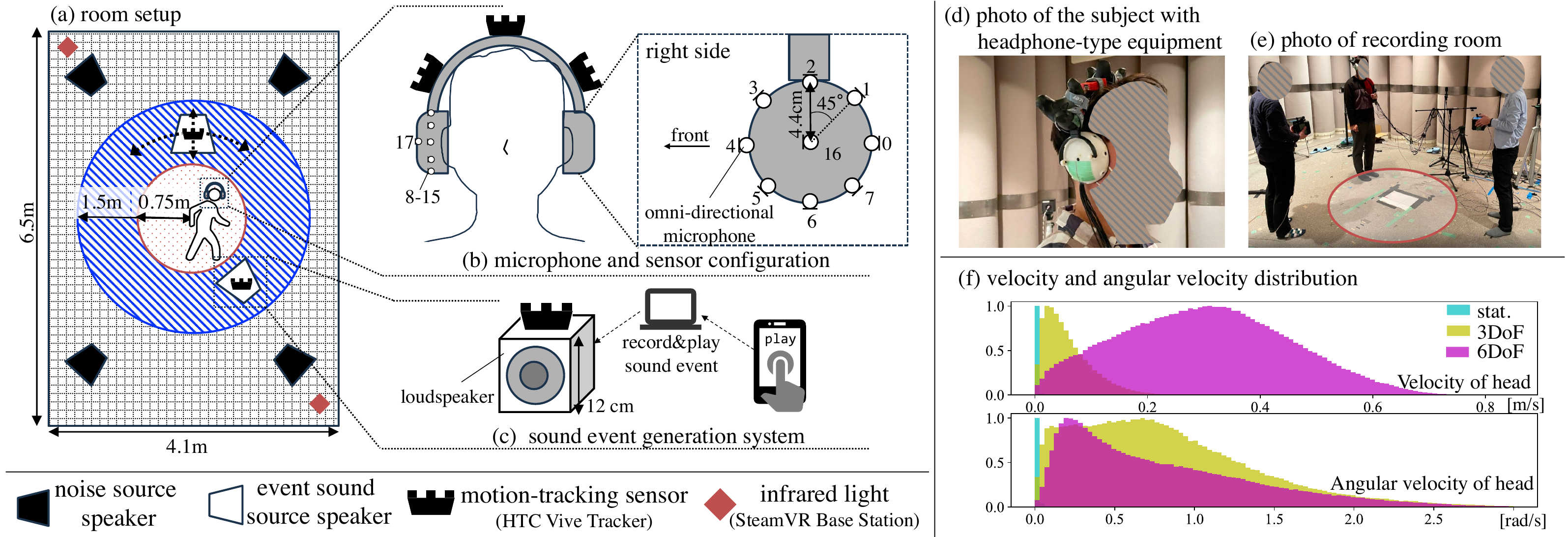}
\vspace{-5pt}
  \caption{Recording setup and equipment configuration for \textit{6DoF SELD Dataset}. In (a) and (e), red circles indicate the range of movement of the subject and blue circles indicate the range of the sound source position.}
  \label{fig:datasetsetup}
  \vspace{-5pt}
 \end{figure*}
 
\begin{table*}[t!]
\centering
\caption{Specification of conventional and proposed SELD datasets.}
\vspace{-8pt}
\scalebox{0.83}[0.83]{
\label{tab:datasetoverview}
\begin{tabular}{@{}l|ccccccccccc@{}}
\toprule
          &Amount  & Event & Other          & Self-motion &       & Source     & \# of & \# of & SNR &   &             \\
Dataset   &of data  & generation & modalities     & of mic.      & Subjects & movement &classes&mic.   & [dB] & Rooms   \\ \midrule
{\footnotesize TAU-NIGENS Spatial} & \multirow{2}{*}{13.3h} & \multirow{2}{*}{IR-based} & \multirow{2}{*}{-} & \multirow{2}{*}{stat.}  & \multirow{2}{*}{-} & \multirow{2}{*}{Moving}  & \multirow{2}{*}{12} & \multirow{2}{*}{4} & \multirow{2}{*}{6 - 30} & \multirow{2}{*}{18}\\
{\footnotesize Sound Events 2021~\cite{dataset2021}}  &  &  & & &  &  &  & & & \\
STARSS23~\cite{SELD2023}  & 7.4h & Real      & Video          & stat.           & -   & Moving   &   13  &4    & natural &    16      \\
Wearable SELD~\cite{WearableSELD}& 8.3h & IR-based & -    & stat.            & 1 (HATS) &  Fixed    &12    & 12 & 10 - 20, clean &  3    \\
Ours: 6DoF SELD &  20.1h & Speaker  & Motion tracker & stat./3DoF/6DoF  & 3 (Human) & Fixed    & 12 & 18& 6 - 20 &   3   \\ \bottomrule
\end{tabular}
}
\vspace{-13pt}
\end{table*}

On the other hand, applications that support moving humans, such as pedestrian safety support, require the ability to handle self-motion microphones. In this problem setting, rapid relative motion of the surrounding sound sources is caused by self-motion, especially rotation. 
For example, given a look-back motion in one second, all surrounding sound sources move at $180$ deg./sec., faster than a $60$ km/h car crossing 3 meters from a human.
Considering that moving sound sources degrade the localization performance in the conventional SELD task~\cite{seldproblem}, this rapid relative motion should also cause degradation. 
Adapting to such rapid system changes is considered more difficult for online systems that use only past observations.

In contrast to the difficulties in 6DoF SELD, it has been reported that humans can localize sound sources more accurately when their heads are moving rather than stationary~\cite{selfmotion1,selfmotion2,selfmotion3}.
The major reason for this is that dynamic cues, such as dynamic changes in inter-time difference and inter-level difference, play an important role in source localization during self-motion~\cite{selfmotion1}.
Considering these human auditory characteristics, utilizing dynamic cues in the 6DoF SELD system is a promising strategy.
For this purpose, training data in the 6DoF SELD situation needs to be collected to train a system that can capture the dynamic cues of acoustic features. 
Furthermore, if the system can observe self-motion in the same way humans use semicircular canals, it is expected to capture dynamic cues more effectively.
In fact, the effectiveness of using head rotation information in binaural source localization has been reported~\cite{headdoa}.
Observation of self-motion is the low-cost option by using inertial sensors, which are commonly used in wearable devices.

Therefore, we propose and publish a new dataset for 6DoF SELD, called \textit{6DoF SELD Dataset}$^1$. Unlike conventional SELD datasets, our dataset is designed to identify sound events that occur around a moving human. It uses headphone-type equipment with three motion tracking sensors and 18-channel microphones to measure the position and posture of the head and acoustic signals.

We also propose a new multi-modal SELD system that combines acoustic signals with velocity and angular velocity observations from motion tracking sensors.
The system introduces sensor signal-based excitation of the acoustic features to mimic humans' ability to utilize dynamic cues. It was implemented by introducing a multi-modal transfer module (MMTM) proposed for multi-modal speech enhancement and action recognition into SELDNet, the baseline model for the DCASE 2023 task3.
Numerical experimental results showed that training the system using the data with the proposed dataset improved 6DoF SELD performance compared to using the data from a stationary microphone array.
In addition, using sensor information of velocity and angular velocity was shown to effectively improve 6DoF SELD performance.

\vspace{-5pt}
\section{Proposed Dataset}
\vspace{-5pt}
In this section, we describe an overview and specifications of our proposed \textit{6DoF SELD Dataset} and a comparison with conventional SELD datasets.
\label{sec:proposedmethod}
\vspace{-5pt}
\subsection{Dataset overview}
\vspace{-3pt}
We propose a \textit{6DoF SELD Dataset}$^1$ for detecting and localizing sound events from the view of a self-motioning human. Unlike conventional SELD datasets with a fixed microphone array~\cite{dataset2019,dataset2020,dataset2021,SELD2023} or a wearable SELD dataset with HATS~\cite{WearableSELD},  we record sound events with headphone-type equipment worn by a  subject with 6DoF self-motion, i.e., walking and looking around.
The 18-channel microphone array and three motion tracking sensors are installed in headphone-type equipment. Motion tracking sensors allow us to observe the position and posture of the head.
By time differentiating the position and posture acquired by the motion tracking sensors, it is also possible to simulate the observation of head motion by a more practical sensor such as a 6-axis inertial measurement units (IMUs).

\vspace{-5pt}
\subsection{Dataset and equipment specifications}
\vspace{-5pt}
Figure~\ref{fig:datasetsetup} shows the recording setup and equipment configuration for the \textit{6DoF SELD Dataset}.
The recording was conducted in a variable reverberation room, as shown in Fig.~\ref{fig:datasetsetup} (a). A human wearing headphone-type equipment moves in the red circle area, and sound events are played randomly in the blue area. Fig.~\ref{fig:datasetsetup} (b) shows the details of the headphone-type equipment. Each left and right plastic earpad has an 8-channel microphone on the outer edge and a 1-channel microphone on the center. In the experimental section of this paper, only microphones 0, 4, 8, and 12 of these channels were used. Future studies could include array processing using two circular microphone arrays or a more realistic setup using only the two central microphones.
Three motion trackers are attached to the headband. The head position and posture are observed as centroid and posture of a triangular composed by these three motion trackers.
Fig.~\ref{fig:datasetsetup} (c) shows the sound event generation system. Sound events are generated by randomly playing audio clips from two loudspeakers. Variations of directions of arrival of sound events are reproduced by manually moving the speakers to various heights and angles. The sound source position is recorded by the motion tracking sensor in the absolute coordinate system of the room and then converted to a relative coordinate system to the central human's head on the basis of the observed head posture information.

\begin{figure*}[t]
\begin{center}
\includegraphics[width=\linewidth]{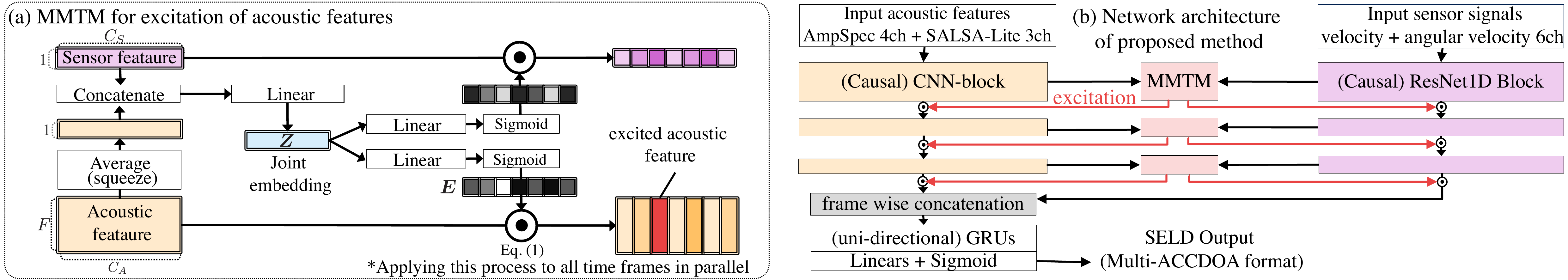}
  \vspace{-15pt}
  
  \caption{(a) MMTM for excitation of acoustic features on the basis of sensor signals. $C_A$ and $C_S$ are the number of channels of the sensor and acoustic features, $F$ is the number of dimensions of the acoustic features, and $\odot$ denotes the Adamar product. (b) Network architecture of proposed multi-modal SELD system. ``AmpSpec'' denotes the amplitude of the spectrogram.}
  \vspace{-24pt}
  \label{fig:architecture}
  \end{center}
 \end{figure*}

Table~\ref{tab:datasetoverview} shows the specifications of the \textit{6DoF SELD Dataset} and the conventional SELD dataset. Our dataset generates sound events by playing back pre-recorded sound samples from the speakers as described above. Impulse response (IR)-based generation has the advantage of increased data volume and a high degree of control over experimental conditions but is unsuitable for self-motion microphones. Using real-life sound events, as in STARSS23, is most compatible with real-life scenarios, but collecting a large amount of labeled data is difficult. Our dataset uses a speaker-based playback of sound events, allowing us to record sound events with the self-motion microphone and collect a sufficient amount of labeled data (20.1 hours).
Our dataset and STARSS23 record multimodal signals as a dataset for SELD. The STARSS23 records 360$^{\circ}$ video of a human generating sound events. Although video modalities can be used to record body movements, they are not suitable for our purpose of SELD with wearable equipment, so we used motion tracking sensors.
The self-motion of the microphone is included only in our dataset. The dataset is divided into three subsets (``stat.'', ``3DoF'', and ``6DoF'') in accordance with the self-motion condition. In ``stat.'' the subject is seated in a chair; in ``3DoF,'' the subject makes rotational movements of the head and body while standing; and in ``6DoF,'' the subject walks, changes direction, and swivels within a circle of $0.75m$. These subsets' actual velocity and acceleration distributions are shown in Fig.~\ref{fig:datasetsetup} (f).
Sound events consisting of 12 classes were recorded by an 18-channel microphone (Fig.~\ref{fig:datasetsetup}-(b)) attached to headphone-type equipment. Signal-to-noise ratios (SNR) were controlled to 6, 10, and 20 dB, allowing quantitative analysis of the system's noise robustness. The recording room is equipped with a variable reverberation room, where the reverberation time is controlled in three steps ($T_{60}^{500{\rm Hz}} = 0.12, 0.30, 0.41$ sec).

All microphones used for recording were Hosiden KUB4225, with a sampling frequency of 48k and a bit depth of 16 bits. A combination of HTC Vive Tracker (2018) and HTC SteamVR Base Station 2.0 was adopted for the motion tracking sensors. The sensor signals were recorded at a nonuniform sampling rate of about 40 fps and then downsampled to a uniform sampling rate of 20 fps.

\vspace{-5pt}

\section{Proposed method}
\vspace{-5pt}
In this section,  we describe the proposed wearable SELD system utilizing the joint feature of audio and sensor signals.
\vspace{-3pt}
\subsection{Basic concept}
\vspace{-3pt}
For a 6DoF SELD system to maintain robust performance during self-motion, it is desirable to exploit appropriate acoustic features as cues depending on the state of self-motion.
Estimating self-posture from acoustic signals has been explored~\cite{CMArot,echo1,echo2}, but online estimation of velocity and angular velocity, which are considered relevant to the dynamic cue, is still difficult.
Utilizing tracking sensor observations, especially velocity and angular velocity, is expected to enable these self-motion states to be acquired and the system to more appropriately utilize dynamic cues to improve SELD performance.

Therefore, we propose a multi-modal SELD system that combines acoustic signals with velocity and angular velocity signals obtained by tracking sensors.
This system mimics the mechanism by which people utilize dynamic cues as excitation of specific acoustic features on the basis of velocity and angular velocity.
It is represented by weighting the $C_A$ channel and $F$ dimensional acoustic features $\bm{\phi}^c\in\mathbb{R}^{F}$ ($c\in[1,C_A]$) by the excitation vector $\bm{E}\in\mathbb{R}^{C_A}$, which depends on velocity $\bm{\nu}\in\mathbb{R}^{3}$ and angular velocity $\bm{\omega}\in\mathbb{R}^3$:
\vspace{-3pt}
\begin{equation}
\label{eq:excitation}
[\tilde{\bm{\phi}}^1,\ldots,\tilde{\bm{\phi}}^C]
= [\bm{\phi}^1,\ldots, \bm{\phi}^C]\odot \bm{E}(\bm{\nu},\bm{\omega})
\vspace{-3pt}
\end{equation}

As an implementation of this principle, we used MMTM~\cite{mmtm}, a method for fusing convolutional neural network (CNN) features of multi-modal signals. 
Fig.~\ref{fig:architecture} (a) shows the block diagram of the MMTM applied to our problem setting. 
In MMTM, the acoustic features obtained at each layer of the CNN are first averaged in the frequency axis. This operation is called ``squeeze'' in the original MMTM. From these squeezed acoustic features and sensor features, a joint embedding $\bm{Z}\in\mathbb{R}^{D_z}$is then extracted.
Finally, The excitation for each modality feature is computed based on $\bm{Z}$. 
Note that in our implementation, the original MMTM is modified to avoid squeezing the time axis of the features to excite the appropriate acoustic features at each time frame.
Since the excitation to acoustic features is dependent on the sensor signal through the joint embedding $\bm{Z}$, MMTM is an appropriate implementation of the principle of Eq.~\ref{eq:excitation}.

\vspace{-6pt}
\subsection{Implementation details}
Fig.~\ref{fig:architecture} (b) shows the network architecture of the proposed method. The inputs to the system are 4-channel acoustic signals and 6-channel velocity/angular velocity signals. The 4-channel acoustic signals are composed of channels 0, 4, 8, and 12 of the 18 microphones in the dataset. It corresponds to the microphones embedded in the front and rear of the left and right earpads. 
From these acoustic signals, 4-channel amplitude spectrograms and 3-channel SALSA-Lite features~\cite{SALSA-Lite} are extracted as input features. The velocity/acceleration signals are extracted by first-order differentiation of the position/angle obtained from the tracking sensor observation signals while smoothing them with a Savitzky-Golay filter~\cite{savgol}.

The DNN used Causal-SELDNet, a causal modification of SELDNet, which is the baseline model of DCASE 2023 task 3~\cite{dcase2023baseline}.
For causal modification, there are three modifications to the network. The first is the use of causal convolution, which uses only past time frames for convolution in the CNN; the second is the change of the bi-directional gated recurrent unit (GRU) to a uni-directional GRU; and third is the removal of multi-head self-attention (MHSA). The DNN for sensor signals was a 1D CNN with ResNet-like skip connections, as in previous studies of action recognition~\cite{mmact}. The 1D CNN was also modified to use causal convolution. 2D and 1D CNNs for extracting features from acoustic and sensor signals consisted of the same number of blocks and time frames, and MMTM excitation was applied to all outputs for each time frame. The acoustic and sensor features obtained as outputs of the CNNs were concatenated in the feature axis and input to the subsequent uni-directional GRUs. Multi-ACCDOA was used as the loss function for training~\cite{multiaccdoa}.

\vspace{-5pt}
\section{Experiments}
\vspace{-5pt}
\subsection{Experimental setup}
\noindent
{\bf Hyper-parameter:} For the short-time Fourier transform, a Hanning window of 1024 points and a shift of 0.025 sec. were used.
From the extracted spectrograms, the frequencies corresponding to 50 - 9050 Hz were cut out, resulting in a frequency dimension of 64.
All model parameters of Causal-SELDNet are the same as in the DCASE2023 baseline model. The 1D CNN used for the encoder of the sensor signal consisted of three ResNet blocks. The number of CNN filters was (64, 32, 16), kernel size, stride, and padding were 5,1,2, respectively.
The acoustic and sensor features extracted by the CNN and 1D CNN were concatenated for each time frame.
The Adam optimizer was used for learning, with an initial learning rate of 0.01~\cite{Adam}. Learning was concluded in 100 epochs, and the parameters obtained in the epoch with the lowest validation loss were adopted.

\vspace{3pt}
\noindent
{\bf Comparison methods:} To evaluate the effectiveness of the proposed method, the following conditions were compared: 
\setlength{\leftmargini}{10pt}    
\begin{description}
\vspace{-2pt}
 \setlength{\parskip}{0cm} 
 \setlength{\itemsep}{0.02cm} 

\item[(A) Baseline (stat.)] A method using Causal-SELDNet trained on the "stat." subset of our dataset. This is a similar situation to the conventional Wearable SELD Dataset~\cite{WearableSELD}, which deals with conditions where the head is fixed. We used this system to investigate the performance degradation of a system trained with fixed microphone data under self-motion conditions.
\item[(B) Baseline] A method using Causal-SELDNet trained on all the data in our dataset. A variant of this method (B/3), trains Causal-SELDNet on 1/3 of the data in the dataset. Since this is the same amount of data as (A), the change in performance can be validated when training data with and without the self-motion microphone under the same amount of data.
\item[(C) Audio-SENet] A method using modified Causal-SELDNet that replaces the CNN with the squeeze and excitation network (SENet)~\cite{senet}. This method can be considered as an uni-modality version of the method (E) using MMTM. Comparing it with (E), the change in performance due to the use of sensor signals can be validated.
\item[(D) Sensor-concat] This method is a variation of (E) that directly concatenates sensor signals and acoustic features extracted using CNN. By comparing this method with (E), the performance improvement by using sensor signals for the excitation of acoustic features based on self-motion can be clarified.
\item[(E) Sensor-MMTM] Proposed method described in Sec.~\ref{sec:proposedmethod}
\end{description}

\vspace{3pt}
\noindent
{\bf Evaluation metrics:} We adopt the same metrics with DCASE 2023 task3~\cite{metrics}.
The metrics used for event detection were location-dependent F1-score $F_{\leq\Theta}$ and error rate $ER_{\leq\Theta}$. These are calculated by counting as true positives (TP) if the event class matches the ground truth label and the event localization is correct within the threshold angle $\Theta$. In this experiment, we adopt $\Theta=20^{\circ}$ as in DCASE 2023 task3.
Class-dependent localization error $LE_{CD}$ and localization recall $LR_{CD}$ were used as metrics for event localization. $LE_{CD}$ is the angular error between the estimated source location and the ground truth, calculated using only TP time frames; $LR_{CD}$ represents recall of the number of active source estimations. 
All experiments were performed three times for different initial parameters. All experimental results are shown with the standard error of the metrics obtained from the three experiments.

\vspace{-4pt}
\subsection{Result}
\vspace{-3pt}
Table~\ref{tb:result1} compares SELD performances with and without self-motion of a subject in the training data.
First, a performance comparison for all test data shows that (B/3) outperforms (A) on all metrics. 
Next, comparing performance under different self-motion conditions in (A), a performance gap exists between the ``stat.'' and ``3DoF'' conditions. 
It suggests that systems trained only on the stationary microphone data cannot adequately cope with the rotation motion.
In addition, the ``3DoF'' subset performs poorly compared to the ``6DoF'' subset. It is considered because ``3DoF'' contains faster rotational motion than ``6DoF'' as shown in Fig.~\ref{fig:datasetsetup} (f).
On the other hand, in (B/3), the performance gap between the ``stat.'' and ``3DoF'' conditions is reduced for the ``3DoF'' and ``6DoF'' conditions. 
These results suggest that 6DoF SELD requires not only a dataset under conventional stationary conditions but also a dataset including the proposed self-motion.

Table~\ref{tb:result2} compares SELD performances for different network architectures and input modalities.
The (D) and (E), which use velocity and angular velocity extracted from head tracking sensors as input features, perform better on all metrics than (B) and (C), which only use audio modality.
The performance improvement in (D) indicates the effectiveness of using sensor features for temporal modeling.
The performance improvement in (E) indicates that MMTM-based excitation of acoustic features based on sensor signals is effective for SELD with self-motion.
This fact is consistent with the property that changes in acoustic features obtained when a human moves his/her head can be used as dynamic cues for source localization.
In (D), where the sensor signal is directly input to the GRU responsible for temporal modeling, a certain performance improvement is also observed compared to (B) and (C).
These results indicate that the use of tracking sensors in 6DoF SELD improves feature extraction and temporal modeling, thus enhancing the performance of SELD.

\begin{table}[t]
\centering
\caption{SELD performance for different DoF of self-motion included in the training data. The ``stat.'' ``3DoF'' and ``6DoF'' indicate that the microphone includes data for stationary, rotating, and rotating/translating cases, respectively.}
\vspace{-9pt}
\label{tb:result1}
\scalebox{0.84}[0.84]{
\begin{tabular}{@{}c|cc|cccc@{}}
\toprule
 & \multicolumn{2}{c|}{motion} & \multicolumn{4}{c}{SELD performance} \\ \cmidrule{2-7}
 & \multicolumn{1}{c}{train} & \multicolumn{1}{c|}{test} & $ER_{\leq20^{\circ}}\downarrow$ & $F_{\leq20^{\circ}}\uparrow$ & $LE_{CD}\downarrow$ & $LR_{CD}\uparrow$ \\ \midrule
\multirow{4}{*}{
\begin{tabular}[c]{@{}c@{}}(A)\\ Baseline \\ (stat.)\end{tabular}} 
& \multicolumn{1}{l}{\multirow{4}{*}{stat.}} & all & $0.63_{\pm0.007}$ &$39.0_{\pm0.5}$ &$25.6_{\pm0.2}$ &$83.3_{\pm0.04}$ \\
 & \multicolumn{1}{c}{} & stat. & $0.48_{\pm0.003}$ &$53.8_{\pm0.5}$ &$19.5_{\pm0.1}$ &$85.8_{\pm0.04}$ \\
 & \multicolumn{1}{c}{} & 3DoF & $0.71_{\pm0.01}$ &$32.8_{\pm0.4}$ &$28.4_{\pm0.2}$ &$82.2_{\pm0.2}$  \\
 & \multicolumn{1}{c}{} & 6DoF & $0.69_{\pm0.02}$ &$31.9_{\pm1.3}$ &$28.5_{\pm0.5}$ &$81.8_{\pm0.1}$  \\ \midrule
\multirow{4}{*}{\begin{tabular}[c]{@{}c@{}}(B/3)\\ Baseline \\ (1/3 data)\end{tabular}} & \multicolumn{1}{c}{\multirow{4}{*}{all}} & all & $0.55_{\pm0.01}$ &$45.7_{\pm0.6}$ &$23.0_{\pm0.1}$ &$84.6_{\pm0.5}$  \\
 & \multicolumn{1}{c}{} & stat. & $0.51_{\pm0.01}$ &$50.6_{\pm1.1}$ &$20.5_{\pm0.3}$ &$86.1_{\pm0.4}$  \\
 & \multicolumn{1}{c}{} & 3DoF & $0.57_{\pm0.008}$ &$44.9_{\pm0.5}$ &$24.2_{\pm0.1}$ &$83.9_{\pm0.3}$  \\
 & \multicolumn{1}{c}{} & 6DoF & $0.59_{\pm0.005}$ &$42.0_{\pm0.5}$ &$24.2_{\pm0.1}$ &$83.6_{\pm0.8}$ \\ \bottomrule
\end{tabular}
}
\vspace{-5pt}
\end{table}

\begin{table}[t]
\centering
\caption{Comparison of SELD performance for different network architectures and input modalities.}
\vspace{-9pt}
\label{tb:result2}
\scalebox{0.89}[0.89]{
\begin{tabular}{@{}l|llll@{}}
\toprule
 & \multicolumn{4}{c}{SELD performance} \\ \cmidrule(l){2-5} 
 & \multicolumn{1}{c}{$ER_{\leq20^{\circ}}\downarrow$} & \multicolumn{1}{c}{$F_{\leq20^{\circ}}\uparrow$} & \multicolumn{1}{c}{$LE_{CD}\downarrow$} & \multicolumn{1}{c}{$LR_{CD}\uparrow$} \\ \midrule
(B) Baseline & $0.55_{\pm0.003}$ &$49.1_{\pm0.2}$ &$21.6_{\pm0.1}$ &$85.2_{\pm0.1}$  \\
(C) Audio-SENet & $0.54_{\pm0.003}$ &$51.1_{\pm0.2}$ &$21.2_{\pm0.2}$ &$\underline{85.9}_{\pm0.1}$   \\ \midrule
(D) Sensor-concat & $\underline{0.53}_{\pm0.005}$ &$\underline{51.4}_{\pm0.4}$ &$\underline{20.9}_{\pm0.1}$ &$85.2_{\pm0.2}$  \\
(E) Sensor-MMTM & ${\bf 0.51}_{\pm0.003}$ &${\bf 54.1}_{\pm0.1}$ &${\bf 20.0}_{\pm0.1}$ &$\bm{86.1}_{\pm0.3}$  \\ \bottomrule
\end{tabular}
}
\vspace{-14pt}
\end{table}

\vspace{-4pt}
\section{Conclusion}
\vspace{-4pt}
We designed a \textit{6DoF SELD Dataset} and proposed a multi-modal sound event localization and detection (SELD) system that combines motion tracking sensor signals with acoustic signals. 
Our dataset provides recordings of acoustic events around a subject moving at 6DoF. The data is captured using a headphone-type device with embedded microphones and motion tracking sensors.
The proposed method utilizes dynamic cues by applying excitations to the acoustic features in accordance with the velocity and angular velocity extracted from the sensor signals. Validation experiments on our dataset showed that learning the system using a dataset that includes self-motion improves SELD performance during movement. Furthermore, it also demonstrated that using sensor signals can improve SELD performance. Therefore, the proposed dataset and system effectively perform SELD on a self-motioning human.

\clearpage
\footnotesize
\bibliographystyle{IEEEbib}
\bibliography{refs}

\end{document}